\documentclass[12pt,preprint]{aastex}
\shorttitle{Molecular Gas in Maffei~2}
\shortauthors{Mason \& Wilson}
\begin{document}

\title{Extended Molecular Gas in the Nearby Starburst Galaxy Maffei~2}

\author{A. M. Mason and C. D. Wilson}
\affil{Physics and Astronomy Department, McMaster University,
    Hamilton, Ontario L8S 4M1 Canada}
\email{wilson@physics.mcmaster.ca}

\begin{abstract}
We present a $9^\prime \times 9^\prime$ fully-sampled map of the 
CO $J=1-0$ emission in the 
nearby starburst galaxy Maffei~2 obtained at the Five College Radio
Astronomy Observatory.  The map reveals previously known strong
CO emission in the central starburst region as well as an
extended asymmetric distribution with 
bright CO lines at the ends of the bar and in a feature at
the north-east edge of the molecular disk.
This northern feature, 
proposed previously to be an interacting companion galaxy, could be a 
dwarf irregular galaxy, although the CO data are also consistent
with the feature being simply an extension of one of the spiral arms.  
We estimate the
total molecular gas mass of Maffei~2 to be $(1.4-1.7) \times 10^9 ~ M_{\odot}$
or $\sim3-4\%$ of its dynamical mass. 
Adopting the recently determined lower value for the CO-to-H$_2$ conversion
factor in the central region, our data lead to the
surprising result that the largest concentrations of molecular gas
in Maffei~2 lie at the bar ends and in the putative dwarf
companion rather than in the central starburst.
A gravitational stability analysis reveals that the extended disk of
Maffei~2 lies above the critical density for star formation; however,
whether the central region is also gravitationally unstable depends
both on the details of the rotation curve and the 
precise value of the CO-to-H$_2$
conversion factor in this region.
\end{abstract}


\section{Introduction}

Maffei~2 is a nearby barred spiral galaxy (Hubble-type: SBb(s)
pec; \citealp{HurtIR93,Buta99}) which is a member of the IC342/Maffei group of
galaxies located in the Zone of Avoidance behind the Galactic plane
\citep*{Krismer95}.  Strong infrared \citep{Rick83, HurtIR93},
Brackett line \citep*{Ho90}, and radio continuum emission
\citep*{Seaq76, Hurt96} indicate that the central region is undergoing a
starburst.  Originally thought to be at a distance of 5 Mpc
\citep{Spin73}, a recent study places Maffei 2 somewhat closer at
$3.47 \pm 0.57$ Mpc (\citet{Finger02}, 
corrected to a true distance modulus of 18.5 for the LMC).  
The star formation rate in Maffei~2 is close to that of
other well-studied, nearby starbursting spirals such as NGC 253 and M83
\citep{Turn94}, but its unfortunate location ($ b = -0^{\circ} 19'$) 
has left it poorly studied compared to its
counterparts.  Though well-mapped in HI \citep{Hurt96} and
infrared \citep{HurtIR93,Buta99}, previous CO
observations have concentrated on the nuclear region, rather than the
galaxy as a whole (CO $J=1-0$: \citealp*{Welia88, Ish89}; higher CO
transitions: \citealp{Sarg85, HurtCO93, Israel03}).  While the
early study of 
\citet*{Rick77} observed several positions along the major axis, the full
two-dimensional distribution of the molecular gas in Maffei~2 has not been
previously observed.

Due to its high visual obscuration, the morphology of Maffei~2 was only 
established fairly
recently.  Originally identified as a
spiral galaxy by \citet{Spin71} and later classified as an Sbc II by
\citet{Spin73}, evidence for a bar was not confirmed until a later
high-resolution HI study \citep*{Hurt88}.  Its Hubble type was revised
by \citet{HurtIR93} to SBb(s) pec after infrared observations revealed
asymmetric spiral arms, misaligned bar halves, and an anomalous third
spiral arm that appeared to lead the galaxy.  This third ``arm'' was
interpretated by \citet{HurtIR93} as a possible tidal tail connecting
Maffei~2 with an interacting dwarf galaxy.  
However, \citet{Buta99} identify this feature as part of a trailing spiral
arm and even identify a counterpart spiral arm on the opposite side of
the galaxy in their deep optical images. 
An HI and 21 cm continuum
study by \citet{Hurt96} showed similar features to the infrared data,
as well as revealing a central hole in the atomic gas distribution.
Differences in the systemic velocity between the central regions and the disk
were interpreted to be indicative of tidal disruption \citep{Hurt96}.

Previous CO observations of the central region of Maffei~2
have shown that the molecular gas is strongly concentrated towards the inner
arcminute of the galaxy \citep{Rick77, Welia88}. Aperture
synthesis observations of this central region have identified features such as
a molecular bar and an expanding nuclear ring \citep{Ish89}.
Although \citet{Rick77} provide an estimate of the total molecular gas mass 
from their major-axis data,
a confirmation of the amount of molecular gas 
from a full two-dimensional map is
necessary to compare Maffei~2 to other nearby starburst galaxies.  The
possibility that Maffei~2 is undergoing tidal disruption adds further
motivation to obtaining a complete molecular gas map of this system.

 In this paper, we present a completely sampled $9' \times 9'$
 map of the $^{12}$CO $J=1-0$ line to study the distribution of
 star-forming regions and the molecular gas content of the inner
disk of Maffei~2.  The
 observations and data reduction are discussed in \S{2}.  The CO
 emission map is presented in \S{3}, and the molecular gas mass is
 calculated and compared to other nearby starburst galaxies in \S{4}.  
We perform a disk stability analysis in \S{5} and the paper is summarized in
 \S{6}.

\section{Observations and Data Reduction}

\indent Maffei~2 was observed with the SEQUOIA (Second Quabbin Optical
Imaging Array) single-sideband instrument at the Five College
Radio Astronomy Observatory (FCRAO) 14~m telescope
(half-power beamwidth $=45.5''$) using the ``on the fly'' (OTF)
mapping mode.  SEQUOIA has 32 pixels arranged in a dual-polarized 4
$\times$ 4 array; however, only 
16 of these pixels are available for extragalactic
or Galactic Center observations which require a wider spectrometer bandwidth of
320 MHz (5 MHz/channel).  The OTF maps were obtained on 2002 February
25 and 26.  Each of the 16 pixels was examined individually to check
for bad baselines; data from one bad pixel were excluded from the
first day's data, and data from three bad pixels were excluded from
the second day's data.  In total, 11 complete maps of Maffei~2 were
obtained covering $9' \times 9'$.  The amount of on-source time varied
with position in the map, but most points had between 3 and 4 hours of
integration time.  Pointing and focus checks were done immediately
before, once during, and immediately following the source observations
on each day.  Typical system temperatures were 500-700 K at an
elevation of $> 60^{\circ}$.  Contamination in the off position
towards R Cas from local absorption  at 0 km s$^{-1}$ and Perseus arm
absorption at $-50$ km s$^{-1}$ was removed from the spectra by
interpolating across adjacent channels.  
The data
were noise-weighted, flat-fielded (divided by the relative beam
efficiency for each horn), and a first-order baseline removed from
each spectrum before being written to a CLASS-format file for further
analysis.

The 11 OTF maps of Maffei~2 were co-added to produce one spectrum 
at each observed location.  All antenna temperatures were converted to
main beam temperatures using the relation $T_{mb} = T_{A}^{*} /
\eta_{mb}$, where $\eta_{mb} = 0.45$ for the FCRAO
telescope\footnote{http://donald.phast.umass.edu/~fcrao/observer/}.  The
parameters of the spectra were found by fitting gaussians to the lines
in CLASS.  The typical rms noise in the reduced spectra was $\sim$
0.025 K at a velocity resolution of 13 km s$^{-1}$ (5 MHz).  In the
center, which contains the strongest emission, the peak brightness
temperature was $T_{mb}= 0.50$ K.  The integrated
intensity, $I_{CO}$, was obtained from the gaussian fit to each spectrum.  The
uncertainty in the integrated intensity is $\Delta I_{CO} =
\sigma ~
\Delta v ~ (N_{\rm{line}})^{1/2} ~
(1+N_{\rm{line}}/N_{\rm{base}})^{1/2}$ \citep{Wilson89}, where $\Delta
v$ is the velocity resolution (13 km s$^{-1}$), $\sigma$ is the rms
noise, $N_{\rm{line}}$ is the number of channels in the line, and
$N_{\rm{base}}$ is the number of channels in the baseline. Values
for $\Delta I_{CO}$ ranged from 
0.4 to 1.4 K km s$^{-1}$ for a range of line widths 
(full width half-maximum) from 13 to
190 km s$^{-1}$; the relative uncertainties for each integrated intensity
measurement range from 2\% to 33\%
with a median value of 7\%.  The peak
value of $I_{CO}$, found in the center spectrum, was $86.6 \pm 1.4 $ K km
s$^{-1}$.  

\section{The Spatial Distribution of Molecular Gas in Maffei~2}

\begin{figure}
\plotone{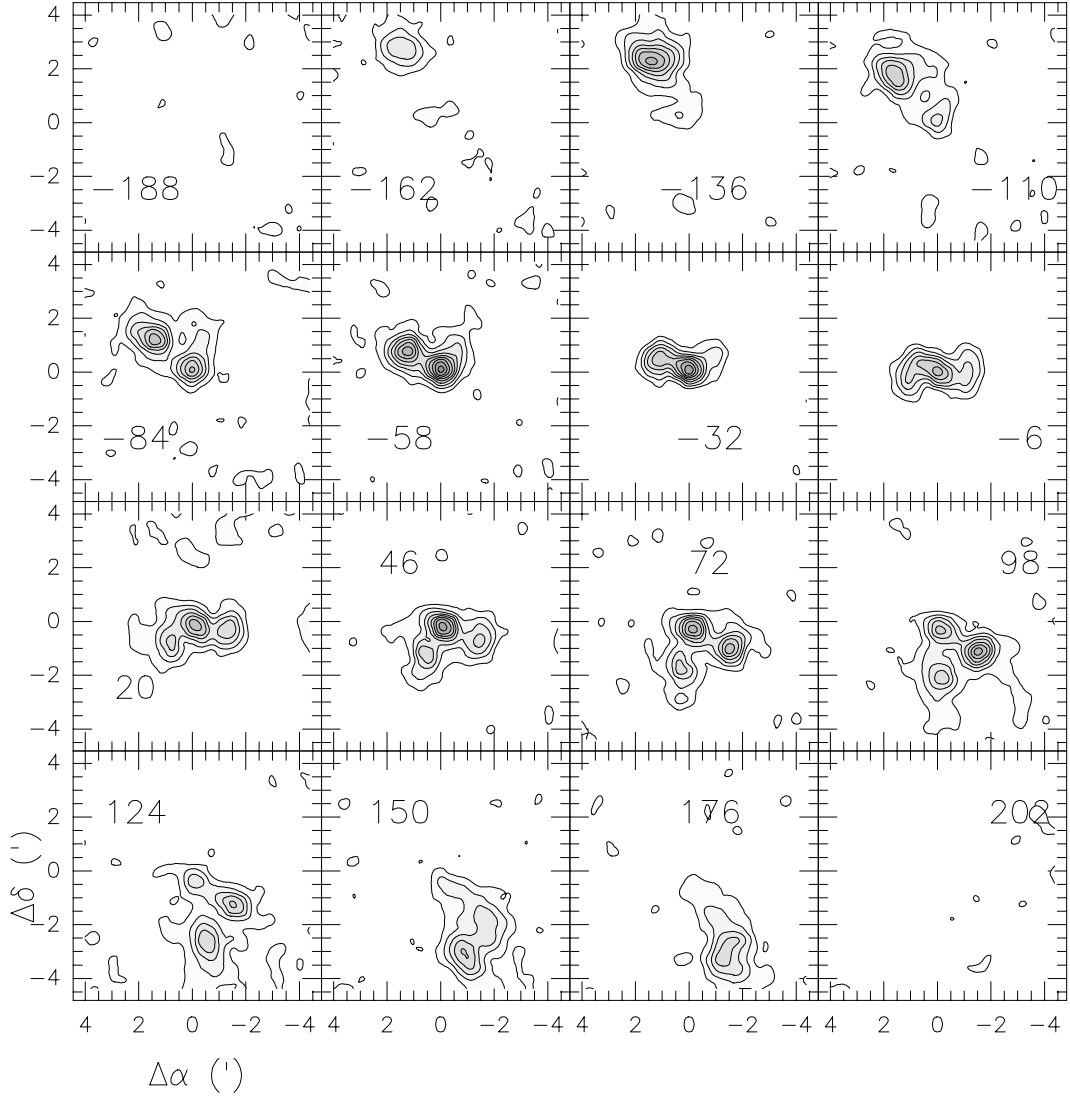}
\caption{Channel maps of CO $J=1-0$ emission from
Maffei~2, labeled by channel velocity $V_{LSR}$ in 
km s$^{-1}$.  The data have been binned by a factor of two to
produce a velocity resolution of 26 km s$^{-1}$ (10 MHz), and a mean
1$\sigma$ noise of 0.018 K. Contour levels are 0.036 K (2$\sigma$) to 0.5 K
by 3$\sigma$ (in T$_{mb}$).  Galactic absorption present in the $-50$
km s$^{-1}$ and 0 km s$^{-1}$ channels was removed by interpolation
from adjacent channels; as a result, the true
emission in those velocity ranges is somewhat uncertain.  The center
coordinates of the map are $\alpha_{2000}$ = 02:41:55.10,
$\delta_{2000}$ = +59:36:18.0.}\label{channelmap}
\end{figure}

\begin{figure}
\plotone{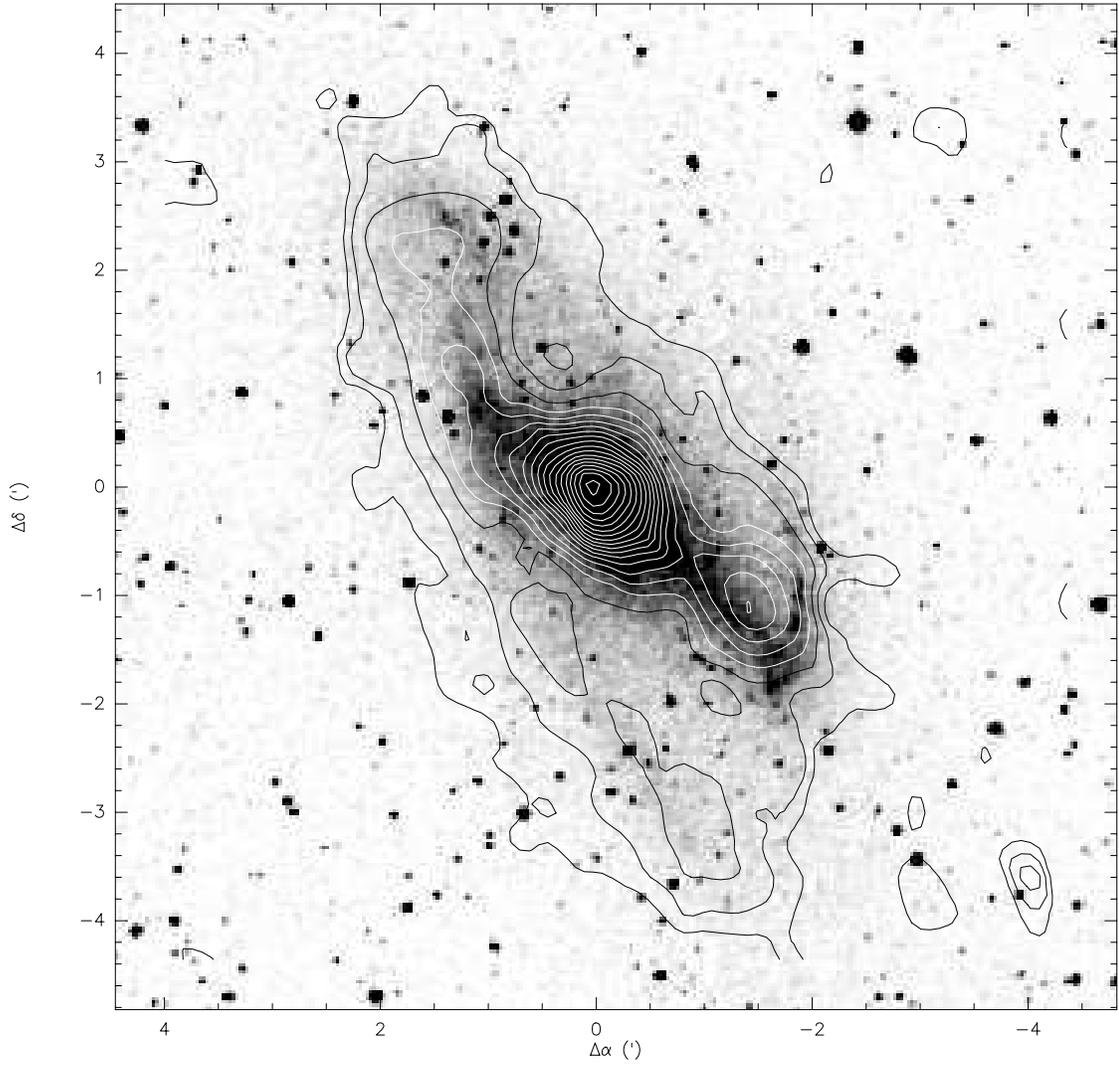}
\caption{CO $J=1-0$ integrated intensity (contours) overlaid on the 2MASS 
K-band image.  Contours start at 4.5 K km s$^{-1}$ and increase by 4.5 K
km s$^{-1}$ (in T$_{mb}$), with the first three contours shown in
black for clarity.  Right ascension and declination offsets
are in arcminutes from $\alpha_{2000}$ = 02:41:55.10,
$\delta_{2000}$ = +59:36:18.0. The infrared and CO images are aligned to within
$3''$.}\label{I_CO_overlay}
\end{figure}

The distribution of the CO emission as a function of velocity
in Maffei~2 is shown in Figure~\ref{channelmap} and 
a contour plot of the integrated CO intensity, $I_{CO}$, overlaid on the 
2MASS K-band image is shown in
Figure~\ref{I_CO_overlay}.  Note that Figure~\ref{I_CO_overlay}
does not show the full extent of Maffei 2, as \citet{Buta99} identified
emission from the galaxy out to a radius of 12$^\prime$. Indeed, even 
the CO extent of Maffei 2 may be somewhat bigger than shown here, as
there are clear detections of CO emission along the southern edge of
the map.
The strongest peak by far is found in the
central region, as expected from previous CO studies.  The prominent
central region contains 20\% of the total CO flux of Maffei~2
within a region $\sim 45''$ (760 pc) in apparent radius.
The presence of a bar could indicate a galactic-wide funneling
effect of molecular gas from the outer parts to the central region,
perhaps fueling the intense starburst while decreasing the amount of
star formation elsewhere \citep{Sak99}.
The galaxy has a strong $I_{CO}$ peak southwest of the nucleus
which is distinct from the central emission.  In comparing the CO map
and the infrared image, it seems likely that this emission originates 
from the end of the bar and the start of a spiral arm.  
To the northeast, the CO emission extends smoothly along a
well-defined bar. The different appearance of the CO emission in
the northeast and southwest sides of the bar is 
the main source of asymmetry in the CO integrated intensity plot; a
difference in the CO intensity in the two bar halves was also noted
by \citet{HurtIR93}.
Because of the high inclination of Maffei~2 and our relatively
low angular resolution,  distinct spiral arms are hard to trace in our CO map.

The northernmost extent of the CO emission raises some interesting 
questions about the morphology of Maffei~2.  There is a bright spot in
the 2MASS image inside the CO peak there, which suggests that the CO
feature is real rather than due to higher noise levels in the outskirts
of the map. The
location of this infrared peak corresponds to features found in
other wavelengths; in both IR \citep{HurtIR93} and 20 cm continuum
\citep{Hurt96}, this object was identified as a possible tidal
feature, perhaps connecting to an interacting companion. Because
there was no evidence for an independent kinematic signature in HI,
the potential companion was proposed to be a dwarf elliptical galaxy,
which would possess little or no atomic gas
\citep{Hurt96}.  
However, the CO emission seen at this position in our map calls
this identification into question, as 
we would not expect a dwarf elliptical to show up
in a survey of molecular gas. 
One possibility is that this potential companion is a dwarf irregular galaxy
whose HI emission is too weak to produce an obviously independent
kinematic signature.  Alternatively, the 
CO, infrared, and 20 cm  emission could arise from an extension of the
spiral arm that starts from the southwest end of the bar (which would
imply that the spiral arms are more similar in
length than previously thought) or could even be part of an outer
spiral arm, as suggested by deep optical images
\citep{Buta99}.   From
a morphological standpoint, it is not obvious which interpretation is
more likely, although the evidence that Maffei~2 exhibits asymmetry
at many wavelengths lends weight to a merger scenario.

\begin{figure}
\plotone{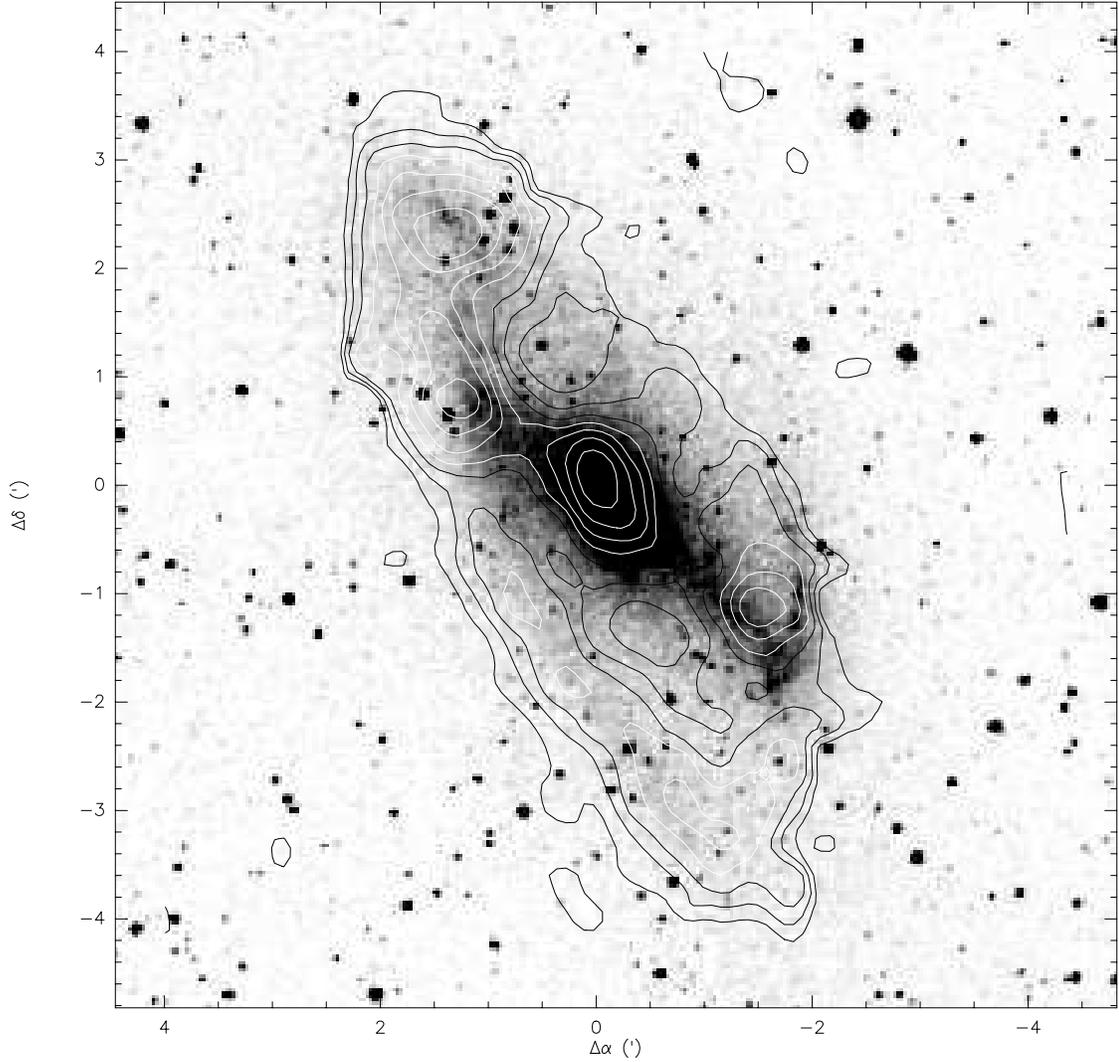}
\caption{Peak brightness temperature, $T_{CO}$, as 
contours overlaid on the 2MASS
K-band image.  The contour levels are 0.10 to 0.40 K by 0.05 K (in T$_{mb}$),
with the first three contours shown in black for clarity.
Right ascension and declination offsets
are in arcminutes from $\alpha_{2000}$ = 02:41:55.10,
$\delta_{2000}$ = +59:36:18.0. The infrared and CO images are aligned to within
$3''$.}\label{T_CO_overlay}
\end{figure}

 A somewhat different picture of the CO distribution can be found 
by comparing the map of peak brightness temperatures, $T_{CO}$, shown
in Figure~\ref{T_CO_overlay}, with the integrated intensity map shown
in Figure~\ref{I_CO_overlay}.  There are several distinct differences
in the plots of $I_{CO}$ and $T_{CO}$.  While the main peaks in the
$I_{CO}$ map are the central region, with a single spectrum integrated
intensity of 86.6 K
km s$^{-1}$, and the southwest bar end, with a single spectrum integrated
intensity 
of 35.8 K km s$^{-1}$, in the $T_{CO}$ map we find four sources with
comparable temperatures: the nucleus, the two bar ends, and the
northernmost peak.  Most of the northeastern emission is negligible
when integrating over velocity, as the lines in this region are
narrow, with typical FWHM of $35-50$ km s$^{-1}$, compared to
$70-85$ km s$^{-1}$ in the southwest.  The galaxy appears less
asymmetric in the $T_{CO}$ map, as the peaks on each side of the
nucleus are at almost the same level.  The spiral arm starting from
the northeast end of the bar is much easier to trace out in the
$T_{CO}$ map, as several bright spots appear to coincide with the
infrared arm.  However, the spiral arm originating from the southwest
end of the bar is still not prominent in the $T_{CO}$ map.

\begin{figure}
\plotone{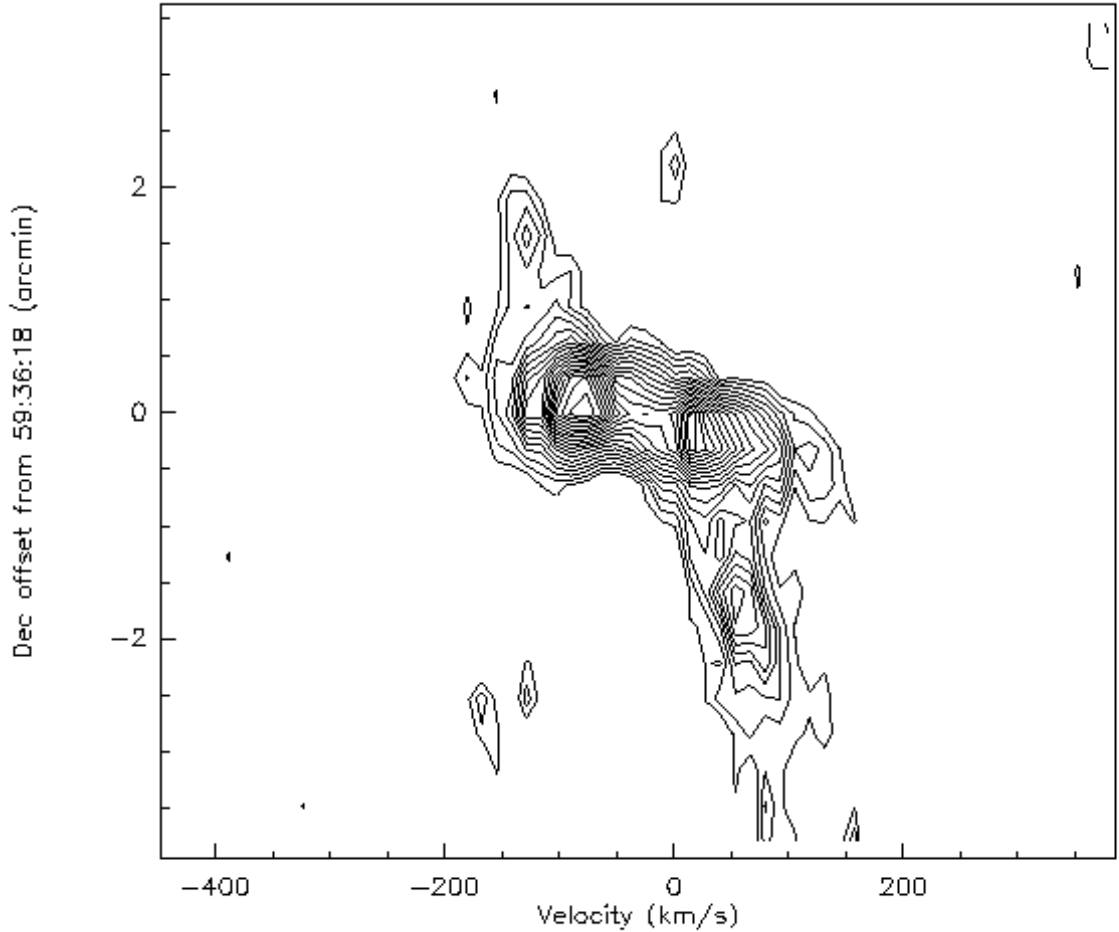}
\caption{Position-velocity map of CO emission along
the major axis of the galaxy (PA=26$^{\circ}$). 
Contour levels are 0.05 to 0.5 by
0.05 K (in T$_{mb}$). Note the double-peaked structure in velocity
space at the origin of the spatial axis, which suggests the
presence of two gas concentrations that are unresolved by
the 45$^{\prime\prime}$ resolution of this data cube.}\label{p-v}
\end{figure}

Although the central emission appears to originate from a
single source in our map, two distinct nuclear components have been
found in other studies with higher angular resolution \citep{Ish89, HurtCO93,
Israel03}.  Surprisingly, even with our large beam, 
we also are able to resolve these two sources
using  a position-velocity cut along the major axis of the galaxy
(Figure~\ref{p-v}).  The two main peaks are separated by
about 100 km s$^{-1}$, although they appear at almost the same
spatial location, which causes line-of-sight superposition in 
Figures~\ref{channelmap}-\ref{T_CO_overlay}. 
The position-velocity slices of the data of \citet{Israel03}
in CO $J=2-1, J=3-2, J=4-3$ show similar peaks separated by 100 km
s$^{-1}$, 80 km s$^{-1}$, and 70 km s$^{-1}$, respectively, with the
change in separation due mostly to increased angular resolution.  We can use
Figure~\ref{p-v} to estimate the systemic velocity of Maffei~2 by finding the
average of the two central velocity peaks.  With the two main peaks
occurring at $-80$ km s$^{-1}$ and 20 km s$^{-1}$, an average of these
gives a systemic velocity 
$V_{LSR} \sim -30$ km s$^{-1}$, which is very close to the
value adopted for most of the CO 
observations\footnote{$V_{LSR}=-30$ km s$^{-1}$ corresponds to
$V_{hel}=-31.8$ km s$^{-1}$, assuming a solar motion of
16.5 km s$^{-1}$ towards  $l=53^\circ$, $ b = 25^\circ$.}
\citep{Ish89, HurtCO93, Israel03}.  An average of the mean velocity at
large spatial offsets gives $V_{LSR} \sim -25$ km s$^{-1}$, which is in
reasonable agreement
with the commonly adopted systemic velocity given our simple analysis.

\section{The Molecular Gas Content of Maffei~2} 

The molecular mass at a single position in our map is given by
$$M_{H_2} = X_{CO} ~ I_{CO} ~ m_{H_2} ~ (1.133 D^2)$$
where $X_{CO}$ is the CO-to-H$_{2}$ conversion factor,
$m_{H_2}$ is the mass of a hydrogen molecule, and the factor $1.133
D^2$ represents the area of a Gaussian beam of full-width
half-maximum diameter $D$.  To find
the mass of regions involving multiple spectra, we must account for
oversampling in the map; this is done by finding the total sum $\sum
I_{CO}$, where each $I_{CO}$ comes from one spectrum, and dividing the
sum by the factor 1.133 $(45''/22'')^2$, which accounts for
the overlapping Gaussian beams.  Using the commonly assumed Galactic value
of $X_{CO} = (3 \pm 1) \times 10^{20}$ cm$^{-2}$ (K km s$^{-1}$)$^{-1}$
\citep{YoSco91}, our previously assumed distance of 3.47 Mpc, and our
summed, sampling-corrected total flux of $540 \pm 10$ K km s$^{-1}$
(where the uncertainty is simply calculated from 
the measurement uncertainties of the individual spectra in the map), we find
the total molecular gas mass of Maffei~2 to be $M_{H_2, tot} = 1.7
\times 10^9 ~ {\rm M_{\odot}}$.  This value is uncertain by about a
factor of 2 due primarily to uncertainties in the CO-to-H$_2$ 
conversion factor.
For comparison, the H$_{2}$ mass of the Milky Way is
estimated to be $2 \times 10^9 ~ {\rm M_{\odot}}$ \citep{ScoSand87}.
Assuming the dynamical mass of Maffei~2 to be $4.7 \times 10^{10} ~
{\rm M_{\odot}}$ \citep{Hurt96}, molecular hydrogen gas
represents $\sim$ 3.6\% of the total mass of Maffei~2.  Adding in the total HI
mass from \citet{Hurt96} and correcting for the new distance, we
obtain a total gas mass $M_{gas} = 2.3\times 10^9 ~ {\rm M_{\odot}}$.
The total hydrogen 
gas content is thus about 4.8\% of the dynamical mass of the galaxy,
surprisingly typical of the values found for SBb galaxies
\citep{YoKne89} given the starburst nature of the central region of Maffei~2.
Note that gas mass to dynamical mass ratios depend linearly on the
rather uncertain value for the distance to Maffei 2.

 It is unlikely that adopting a standard conversion factor for
the entire disk of 
Maffei~2 will result in an accurate assessment of the molecular
content.   Many studies have found that the CO-to-H$_2$
factor can be lower in the nuclei of galaxies by up to a factor of 10 from
the value found in the disk (our Galaxy: \citealp{Sod95, Oka98,
Dahmen98}; other galaxies: \citealp{Israel01, Israel03}).  A recent
paper by
\citet{Israel03} describes a similar result for the nucleus of Maffei
2 and derives a value of $X_{CO} = 2-3 \times 10^{19} ~ {\rm cm^{-2} ~
(K ~ km ~ s^{-1})^{-1}}$.  Their analysis indicates that we may 
overestimate the amount of gas in the central region by about a
factor of ten if we assume a Galactic conversion factor across the
galaxy.  Therefore, we adopt their value $X_{CO} = 3 \times 10^{19} ~ {\rm
cm^{-2} ~ (K ~ km ~ s^{-1})^{-1}}$ to calculate the amount of H$_2$ in
the central region.  With this assumption, we 
find a central molecular mass of $M_{H_2, nuc}
= 3.3 \times 10^7 ~ {\rm M_{\odot}}$ out to a radius of about 760 pc.
We assume that a constant conversion factor across the disk will give
a good estimate of the amount of H$_{2}$, and use the Galactic factor
to find the molecular mass outside of the nucleus.  Combining the
results from these two regions, we find a revised total molecular mass
of $M_{H_2} = 1.4 \times 10^9 ~ {\rm M_{\odot}}$.  This value is still
well within uncertainties of our first estimate; the significant
difference comes in our analysis of the gas content of the nucleus.
\citet{Hurt91} estimated the dynamical mass inside the central R$ <
350$ pc to be $2.8 \times 10^8 ~ {\rm M_{\odot}}$; our value for 
the molecular gas mass in a
region of approximately the same size, using the smaller conversion
factor, is $2 \times 10^7 ~ {\rm M_{\odot}}$.  Without using a
modified value for $X_{CO}$, our observations would lead to $M_{gas} \simeq
 M_{dyn}$ in the central region, an unphysical result which does not reflect
the presence of stars in the galactic center.

The nucleus of Maffei~2 has been studied well in CO, and
provides a good check to our data.  We first convert the values of
 previous authors to the new distance estimate of 3.47 Mpc and the
 nuclear conversion factor found by \citet{Israel03}.
\citet{Ish89} found the nuclear molecular bar (a region 700 pc
$\times$ 140 pc in size) to have a mass of $1.5 \times 10^7 ~
{\rm M_{\odot}}$, and \citet{Welia88} found the molecular mass inside
the central 550 pc, a significantly smaller region than ours, to be
$7.2 \times ~ 10^6 ~ {\rm M_{\odot}}$.  Our higher mass value is
likely due to the larger region we studied.  An interesting result is
found by comparing our data to the data of \citet{Rick77}.  Although
they used a larger beam than ours ($65''$ versus our $45''$), the
peak $I_{CO}$ values (in units of T$_{A}^*$) for the nucleus are almost
the same.  For a point source, we would expect the 
intensity measured by
their beam to be $(45''/65'')^2$ times our value.  The fact that their
intensity is comparable to ours means that either the nuclear emission does
not arise from a point source, or emission from the disk is
significant, or both.  It has been confirmed in several studies that the
nuclear emission arises from two distinct components
\citep{Ish89, HurtCO93, Israel03}, and our position-velocity slice
shows similar structure (see Figure~\ref{p-v}).  Larger-scale high
resolution studies would be necessary, however, to determine the
relative contributions of emission from the nucleus and the
surrounding disk.

The strong off-center flux peak located southwest of the nucleus 
has a molecular mass of $M_{H_2} = 1.6 \times 10^8 ~ {\rm M_{\odot}}$
(using the Galactic conversion factor), which is $\sim 10$\% of the
total molecular mass.  We found the virial mass of this region using
the formula $M_{vir} = 99 \Delta v^2D$, where $\Delta v$ is the
velocity width of the line in km s$^{-1}$, and $D = 800$ pc is the deconvolved
diameter of the area in pc.  Our value of $M_{vir} = 1 \times 10^9
~ {\rm M_{\odot}}$ indicates that we are not observing a gravitationally
bound region.  The final region we studied was the possible
interacting companion galaxy, seen as the northernmost peak
of emission in the temperature contour plot 
(Figure~\ref{T_CO_overlay}).  The amount of flux in that area
corresponds to a mass of $M_{H_2} = 8.8 \times ~ 10^7 ~ {\rm
M_{\odot}}$.  The virial mass of this region (with $D=1300$ pc), 
$M_{vir} = 3.7 \times
10^8 ~ {\rm M_{\odot}}$, is again significantly higher than the
molecular gas mass.  Considering our large beam, it is not surprising
that we cannot resolve gravitationally bound objects.  One
surprising result, however, is that if the CO to H$_2$ conversion factor is
indeed significantly 
lower in the central region than the disk, then these off-center peaks have
much larger concentrations of molecular gas than does the central starburst.
If the end of the bar in Maffei~2 is subject to shock
heating, then the standard CO-to-H$_2$ conversion factor may also
overestimate the molecular gas in this region
\citep*{dss86}. However, if the northern-most
feature is in fact a dwarf irregular galaxy, it would likely have
a sub-solar metallicity, in which case the standard conversion factor
would actually result in an {\it underestimate} of the molecular gas
content in this region \citep{wilson95}.

Table~\ref{comp} compares the
properties of Maffei~2 with two other nearby starburst spiral
galaxies, M83 and IC342. 
We compare published results with our
values calculated using the Galactic value of $X_{CO}$ throughout 
Maffei~2, as this is
the procedure that most authors use in their analysis.  We also
recalculate molecular gas mass estimates for M83 and IC342 
using a common CO-to-H$_2$ conversion factor.

M83, a nearly face-on SBc spiral with a pronounced bar, is
somewhat larger than Maffei~2 in dynamical mass \citep{Cros02} and
has a comparable star-formation rate \citep{Turn94}.  It was recently
observed in the CO $J=1-0$ and CO $J=2-1$ lines using the OTF
method at the NRAO 12 m telescope \citep{Cros02}.  
The total mass of molecular gas found for M83 is
$M_{H_{2}} = 3.8 \times 10^9 ~ {\rm M_{\odot}}$, or about 5\% of its
dynamical mass, which is comparable to the value of 4\% for Maffei~2.
The two galaxies also follow the trend
that later-type galaxies have a higher total gas content \citep{YoKne89}:
Maffei~2 has 5\% of its dynamical mass in gas form, while M83 has an
impressive 14\%.
However, the relative contribution of the different phases of the 
interstellar medium in M83 is quite different from Maffei 2; the ratio of
molecular gas to atomic by mass (a distance-independent ratio) 
is $M_{H_2}/M_{HI}$ = 0.60 for M83,
whereas Maffei~2 has a much higher value of 3.1.  Although this is
consistent with the findings that early-type spiral galaxies have
a larger percentage of gas in molecular form \citep{YoKne89}, the
difference does seem extreme.  We note that the observed HI extent
of Maffei 2 is slightly less than the optical extent in the $I$
band \citep{Hurt96,Buta99}, which is unusual for spiral galaxies that are
not in massive galaxy clusters. Perhaps extended, low surface
brightness HI emission is present in the outskirts of Maffei 2, where it would
be difficult to separate from foreground Galactic HI emission. If so,
this additional HI emission would act to reduce the 
$M_{H_2}/M_{HI}$ ratio in Maffei 2.

IC 342, another nearby starburst of Hubble type Scd, has also
been recently observed in CO \citep{Cros01}.  Adopting 3.3 Mpc for
the distance to IC 342 \citep{Saha02},
the dynamical mass of
IC342 is  $\sim 1.7 \times 10^{11} ~ {\rm M_{\odot}}$
\citep{Cros01}.  It has about twice as much atomic gas as molecular
gas, resembling more closely the interstellar medium 
of M83 than Maffei~2.  
Although the latest-type spiral of the three galaxies, 
IC342 does not have the largest
fraction of gas to overall mass, which is unexpected compared to previous
results \citep{YoKne89}.  
More of
the available H$_2$ is found in the central regions of Maffei~2 and M83,
although all three galaxies have distinct molecular bars.  

\section{Implications for Star Formation in Maffei~2}

We have followed \citet{KennSFL89} in analyzing the stability
of the gas disk at different annular radii.  Assuming a thin
isothermal gas disk, the critical gas surface density is given by
\citep{Toomre64, Cowie81}
$$\Sigma_c = \alpha ~ \frac{\kappa c}{3.36 ~ G}$$
where $c$ is the velocity dispersion of the gas, $\kappa$ is 
the epicyclic frequency, and $\alpha$ is the stability constant.  We
adopt the values $c = 6 ~ {\rm km ~ s^{-1}}$ and $\alpha = 0.7$ as in
the analysis of
\citet{KennSFL89}. \citet{Wang94} have noted that the presence of a stellar
disk increases the stability of the gas. Using their formalism,
\citet{Martin01} calculate that, where the stellar survace density is
high, this formula may overestimate $\Sigma_c$ by up to a factor of three.

\indent We calculated $\kappa$ using the form of V(r) calculated in 
\citet{Hurt96} and substituting into the relation
$$\kappa = 1.41 \frac{V}{r} \left( 1 + \frac{r}{V} \frac{dV}{dr} \right)
^{1/2}
$$
 The velocity fit for the HI rotation curve is
$$
V(r) = 3^3 \frac{V_{\rm max} ~ (r/R_{\rm max})}{(1+2\sqrt{r/R_{\rm
max}})~^3}
$$
 where $V_{\rm max} = 172.0$ km s$^{-1}$ at a radius of
$R_{\rm max} = 322''$. (Our form differs from the one in
\citet{Hurt96} by a factor of 3 to let $V(R_{\rm max}) = V_{\rm max}$
and to agree with the original form in \citet{Brandt65}.)  The final
equation we derived for the critical density, assuming a distance of
3.47 Mpc, is
$$
\Sigma_c = 351 ~ \frac{(2+\sqrt{r/R_{\rm max}})~^{1/2}}{(1+2
\sqrt{r/R_{\rm max}})~^{7/2}} ~ {\rm M_{\odot} ~ pc^{-2}}$$

 We used the HI data from \citet{Hurt96} to find
the HI column density at different annuli (Figure 6 of their paper).
$N_{H_2}$ was found by averaging the CO flux at elliptical annuli,
using the parameters for inclination and position angle commonly
assumed for Maffei~2 ($i = 67^{\circ}$; PA $= 26^{\circ}$).  We added
the molecular column density to $N_{HI}$ to find a total column
density N$_{gas} = 1.36 (N_{HI} + 2N_{H_2})$, where the factor 1.36
adds in the contribution from helium.  In our analysis, we 
initially use the standard
Galactic conversion factor for each radius; we subsequently investigate the
effect of adopting a smaller conversion
factor in the central starburst region.

\begin{figure}
\plotone{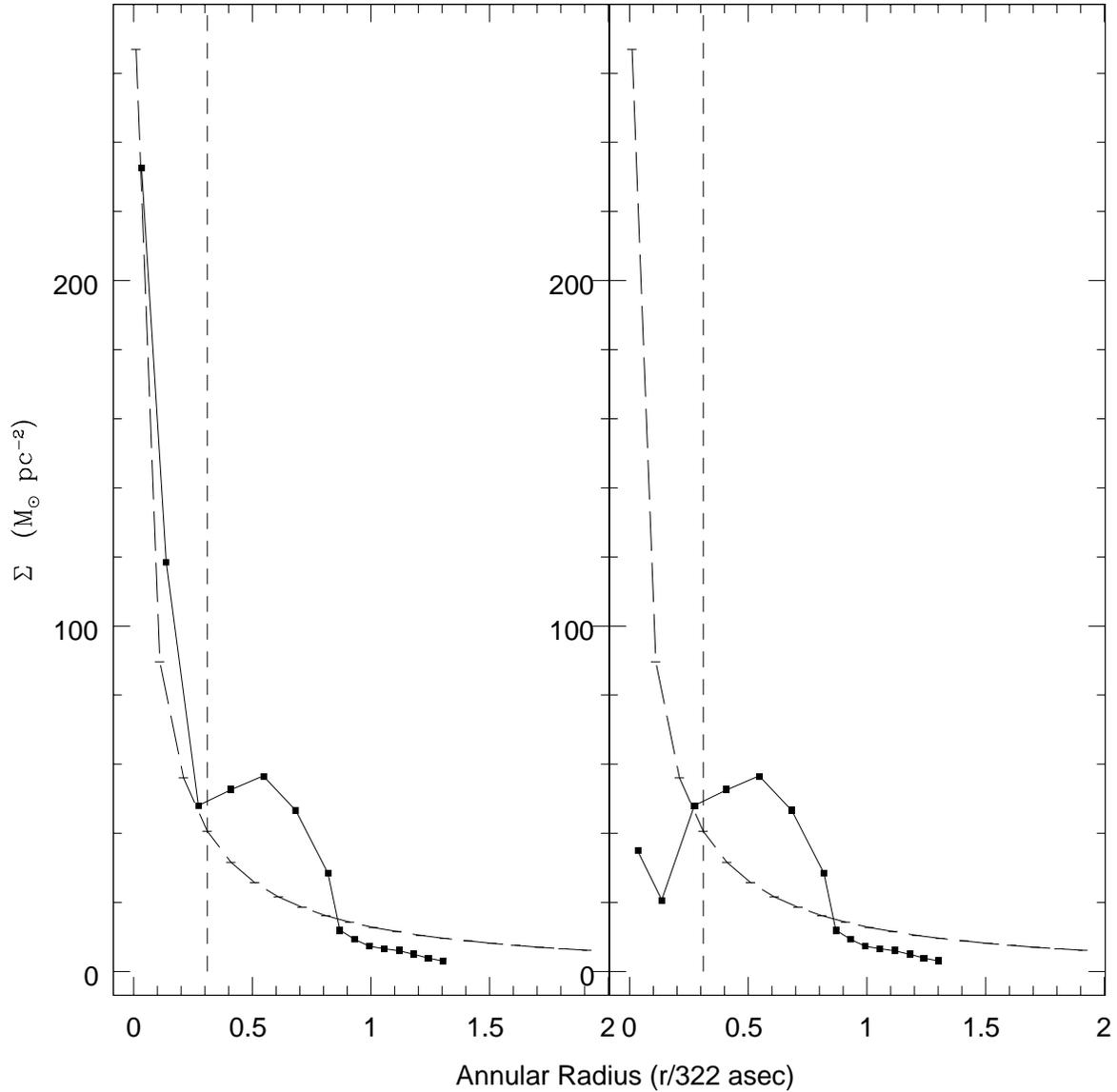}
\caption{(a) The total gas surface density using the standard CO-to-H$_2$
conversion factor as a function of radius in Maffei~2
is shown as a solid line with points.  The critical gas
density for a purely gaseous disk 
is plotted as a dashed curved line; this curve likely overestimates
the true critical density by ignoring the effect of the stellar
disk (see text).  The vertical dashed line
indicates the radius inside which the rotation curve is uncertain
\citep{Hurt96}. (b) The total gas surface density recalculated using
the value $X_{CO} = 3 \times 10^{19} ~ {\rm cm^{-2} ~(K ~ km ~
s^{-1})^{-1}}$ inside a radius of $45''$.}\label{disk}
\end{figure}

 Figure~\ref{disk}a shows a plot of the total gas surface
density $\Sigma_{gas}$ and the critical density $\Sigma_c$ versus
radius.  In the regions where $\Sigma_{gas} > \Sigma_c$, the disk
should be unstable to gravitational perturbations and is likely to be
actively forming stars.  $\Sigma_{gas}$ lies above the critical limit
at every radius that CO is observed, out to about $280''$, or 4.7 kpc.
The ``hump'' in the observed gas density curve between about $88''$
and $280''$ represents the region where the majority of the HI resides
in a large ring outside of the central region. Molecular
emission is negligible outside  $280''$ (4.7 kpc), where HI becomes the
dominant contributor to the total gas surface density; as $\Sigma_{gas}$ dips
below the critical limit in this area, the disk should be stable and
we would not expect to find much star formation.  It is interesting
that our values of $\Sigma_{gas}$ all lie very close to the critical
limit, to within a factor of $\sim 2.5$, a tendency which has been
found for other galactic disks \citep{KennSFL89}.  Although the model
we have used in calculating $\Sigma_c$ is somewhat simplified, it
seems to reproduce the observed gas surface density quite well.

The new result for a modified conversion factor in the central region
leads to a slightly different conclusion (see Figure~\ref{disk}b).
The gas density in the center appears to be below the critical
density, thus indicating that star formation should  happen mainly
outside the central region.  This result clearly does not fit well with the
observations that indicate Maffei~2 is undergoing a nuclear starburst.
However, the rotation curve calculated by \citet{Hurt96} was noted to
be uncertain inside a radius of 100$''$, which encompasses the
majority of molecular gas, including the nucleus.  Although it may be
hard to fit a rotation curve to the central regions of the galaxy, where
non-circular orbits may be dominant, a modified velocity fit in the
center could rectify the stability analysis in this region.  It is
also possible that the central region is at a different inclination than the
disk of the galaxy; such a warp could change the rotation curve enough
to fit the gas density profile. Finally, the central region 
has the highest stellar surface density and thus $\Sigma_c$ is most
likely to be overestimated in this region given our simple analysis
\citep{Wang94}.

However, we can  explore the central dynamics somewhat further
using recent observations.  \citet{Israel03} use their high excitation
CO observations to estimate the velocity gradient in the central
region of Maffei~2.  Correcting the observed value of 18 km s$^{-1} /
''$ for the inclination of the galaxy, we find $dv/dR = 1200$ km
s$^{-1}/$kpc.  Assuming that the rotation curve is linear, the
critical density in the inner $5''$ radius would be $\Sigma_c = 700 ~ {\rm
M_{\odot} ~ pc^{-2}}$; including the effect of stars would 
decrease the value of $\Sigma_c$ by perhaps a factor of three
\citep{Martin01}.  This critical density 
is much larger than the average gas
surface density we find in the central region.  
However, we note that the
molecular gas in the central region appears to reside in two
kinematically-distinct peaks rather than in a single 
central source.  One possible explanation for the observed central
starburst is that there is a change in the rotation curve just past
$5''$, where the gas collects and becomes gravitationally unstable.
Such ``twin peaks'' have been found in other barred galaxies, and can
be explained by orbit crowding near inner Lindblad resonances
\citep{Kenney92}. Even if there is no change in the rotation curve,
the gas could attain sufficient surface densities to be unstable if
it is concentrated into a few compact regions within this region. Such
a model could easily be tested with existing millimeter interferometers.

\section{Summary}  

A complete CO $J=1-0$ map of Maffei~2 reveals a total molecular mass
of $(1.4-1.7) \times 10^9 ~ {\rm M_{\odot}}$.  The central starburst
region produces
a large fraction of the CO flux, but may not contain a similar
portion of the molecular gas mass if the CO-to-H$_2$ conversion
factor is smaller in this region \citep{Israel03}.  The CO
integrated intensity map is asymmetric, with the south-west portion of
the bar much brighter than the north-east half; however, the
distribution of the peak brightness temperature is much more symmetric.
If the smaller conversion factor from the study of \citet{Israel03} is
correct, our analysis leads to the surprising result that three CO
peaks in the outer disk of Maffei~2 each contain more molecular gas 
than does the central starburst.
A global comparison with two other nearby starburst galaxies reveals
that Maffei~2 is rich in molecular gas, although its total gas content
(atomic plus molecular) accounts for 
a smaller fraction of the dynamical mass in Maffei~2 than in M83 or
IC 342.

We have examined the dynamical stability of the gas disk following the method
in \citet{KennSFL89}. The results of the analysis depend heavily on
the value of the CO-to-H$_2$ conversion factor adopted for the central
region.
If we adopt the standard Galactic conversion factor 
across the disk of Maffei~2, then the gas surface density is greater than the
critical density throughout the molecular disk.  On the other hand, if
a smaller value for the conversion factor is appropriate for the 
starburst nucleus, then the gas disk is found to be dynamically unstable
and hence prone to star formation only 
in the large HI ring which lies $100'' - 300''$ from the center of the
galaxy. However, we note that the rotation curve is poorly determined
precisely for the central regions where the conflict between the stability
analysis and the observed starburst is most apparent.

Both the CO integrated intensity and peak brightness temperature maps
reveal extended emission to the north of Maffei~2 in the region of
a possible tidal feature first identified in the infrared by \citet{HurtIR93}.
The presence of CO emission suggests that this putative companion is
more likely to be a dwarf irregular rather than a dwarf elliptical
galaxy as suggested by \citet{Hurt96}. Although the CO data are also
consistent with this emission originating in an extended spiral arm,
the asymmetry observed in Maffei~2 at a variety of wavelengths lends
support to the merger hypothesis.

\acknowledgements
This publication makes use of data products from the 
Two Micron All Sky Survey, which is a joint project of the University
of Massachusetts and the Infrared Processing and Analysis Center,
funded by the National Aeronautics and Space Administration and the
National Science Foundation. We acknowledge the use of NASA's 
{\it SkyView} facility
(http://skyview.gsfc.nasa.gov) located at NASA Goddard
Space Flight Center. The authors wish to thank the referee, Marshall
McCall, for comments which substantially improved this paper.
This research was supported through grants to C. Wilson from the Natural
Sciences and Engineering Research Council of Canada.

\newpage

\begin{table}\footnotesize
\begin{center}
\begin{tabular}{c c c c}\hline
Property                             & Maffei~2             & M83                
& IC342 \\ \hline \hline
Hubble-type                          & SBb                  & SBc                
& Scd\\
Distance (Mpc)                       & 3.47                  & 4                
& 3.3\\
$M_{\rm dyn} ~ {\rm (M_{\odot})}$    & $4.7 \times 10^{10}$ & $7 \times 10^{10}$ 
& $1.7 \times 10^{11}$\\
$M_{\rm H_2} ~ {(\rm M_{\odot})}$    & $1.7 \times 10^9$    & $3.8 \times 10^9$  
& $3.0 \times 10^9$\\
$M_{\rm H_2}/M_{\rm H_I}$            & 3.1                  & 0.6                
& 0.5\\
$M_{\rm H_2, nuc}/M_{\rm H_2, tot}$  & 0.1                  & 0.1                
& 0.05\\
$M_{\rm H_2}/M_{\rm dyn}$            & 0.04                 & 0.05               
& 0.02\\ 
$M_{\rm gas}/M_{\rm dyn}$            & 0.05                 & 0.14               
& 0.05\\
\hline
\end{tabular}
\end{center}
\caption[Comparison: properties of Maffei~2, M83, and IC342]
{The general properties of Maffei~2, M83, and IC342 presented for
comparison.  Values for M83 are taken from \citet{Cros02}, and values
for IC342 are taken from \citet{Cros01}.  We have adjusted the mass
estimates for our choice of conversion factor ($X_{CO} = 3 \times
10^{20} ~ {\rm cm^{-2} ~ (K ~ km ~ s^{-1})^{-1}}$).
Dynamical masses are total dynamical masses derived from a Brandt model.
$M_{\rm H_2, nuc}$ is taken to be the
molecular mass within the central observed spectrum.  Although our
beam is slightly smaller than the one used in the data of
\citet{Cros01} and \citet{Cros02} ($45''$ vs. their $55''$), it is the
closest comparison we can make with the size of the region they
studied.  The only entry that would be affected by this is the value
of $M_{\rm H_2, nuc}/M_{\rm H_2, tot}$; including a larger region
would scale up the nuclear mass in Maffei~2 and make the fraction
larger. }\label{comp}
\end{table}

\end{document}